\documentclass[aps,prb,amsmath,twocolumn,amssymb,floatfix,showpacs,nofootinbib,longbibliography]{revtex4-2}
\usepackage{MnSymbol}
\usepackage{braket}
\usepackage[dvipsnames]{xcolor}
\usepackage{float}
\usepackage{subfigure}
\usepackage[colorlinks=true,linktoc=page,linkcolor=blue,citecolor=blue,urlcolor=blue]{hyperref}
\usepackage{multirow}
\usepackage[T1]{fontenc}
\usepackage[utf8]{inputenc}
\usepackage[normalem]{ulem}
\usepackage{sidecap,tikz}

\newcommand{\vect}[1]{\boldsymbol{\mathrm{#1}}}
\newcommand{\Eq}[1]{Eq.~(\ref{#1})}
\mathchardef\mhyphen="2D 

\newcommand{\non}{\nonumber}  
\definecolor{lime}{HTML}{A6CE39}
\DeclareRobustCommand{\orcidicon}{\hspace{-1.0mm}
    \begin{tikzpicture}
        \draw[lime, fill=lime] (0.0,0.0) 
        circle [radius=0.15] 
        node[white] {{\fontfamily{qag}\selectfont \tiny \,ID}};
        \draw[white, fill=white] (-0.0525,0.095) 
        circle [radius=0.007];
    \end{tikzpicture}
    \hspace{-3.0mm}
}
\foreach \x in {A, ..., Z}{\expandafter\xdef\csname orcid\x\endcsname{\noexpand\href{https://orcid.org/\csname orcidauthor\x\endcsname}{\noexpand\orcidicon}}
}

\allowdisplaybreaks

\def\-{\sout}

\begin{document}
\title{Local and energy-resolved topological invariants for Floquet systems}  

\author{Arnob Kumar Ghosh\orcidA{}}
\email{arnob.ghosh@physics.uu.se}
\affiliation{Department of Physics and Astronomy, Uppsala University, Box 516, 75120 Uppsala, Sweden}

\author{Rodrigo Arouca\orcidB{}}
\email{rodrigo.arouca@physics.uu.se}
\affiliation{Department of Physics and Astronomy, Uppsala University, Box 516, 75120 Uppsala, Sweden}

\author{Annica M. Black-Schaffer\orcidC{}}
\email{annica.black-schaffer@physics.uu.se}
\affiliation{Department of Physics and Astronomy, Uppsala University, Box 516, 75120 Uppsala, Sweden}

\begin{abstract}
Periodically driven systems offer a perfect breeding ground for out-of-equilibrium engineering of topological boundary states at zero energy ($0$-mode), as well as finite energy ($\pi$-mode), with the latter having no static analog. The Floquet operator and the effective Floquet Hamiltonian, which encapsulate the stroboscopic features of the driven system, capture both spectral and localization properties of the $0$- and $\pi$-modes but sometimes fail to provide complete topological characterization, especially when $0$- and $\pi$-modes coexist. In this work, we utilize the spectral localizer, a powerful local probe that can provide numerically efficient, spatially local, and energy-resolved topological characterization. In particular, we apply the spectral localizer to the effective Floquet Hamiltonian for driven one- and two-dimensional topological systems with no or limited symmetries and are able to assign topological invariants, or local markers, that characterize the $0$- and the $\pi$-boundary modes individually and unambiguously. Due to the spatial resolution, we also demonstrate that the extracted topological invariants are suitable for studying driven disordered systems and can even capture disorder-induced phase transitions. 
\end{abstract}

\maketitle

\section{Introduction} \label{introduction}
Incorporating time-periodic driving not only transmutes a system to out-of-equilibrium but also provides a control knob to engineer non-trivial topological phenomena in what may otherwise be trivial systems~\cite{Kitagawacharacterization2010,kitagawa11transport,lindner11floquet,FloquetGuPRL2011,Rudner2013,Usaj2014,Eckardt2017,oka2019,NHLindner2020,Bao2022,GhoshJPCMReview2024}. In a periodically driven system the energy and momenta satisfy the Floquet-Bloch theorem, which restricts the energy to be in the Floquet Brillouin zone~(FBZ) with a period equal to the driving frequency $\Omega$~\cite{Eckardt2017,oka2019}. The resulting folded energy, or quasienergy in Floquet systems, opens up an interesting avenue to host topologically protected boundary modes; not just at $E=0$ as in static systems, but also at the FBZ boundary at quasienergy $E=\Omega/2$~\cite{Kitagawacharacterization2010,JiangColdAtomPRL2011,Rudner2013,Piskunow2014,Usaj2014,Yan2017,Eckardt2017,NHLindner2020}. Such boundary modes at finite quasienergy are also referred to as $\pi$-modes and do not possess any static counterparts. In addition to recent rapid theoretical developments, topology in Floquet systems has also been demonstrated in several different experimental setups~\cite{WangScience2013,Jotzu2014,McIver2020,Wintersperger2020}.

Although driven systems have become an important venue for exotic topological phases, their topological characterization still offers considerable, but sometimes subtle, challenges. The stroboscopic Floquet operator and the corresponding effective Floquet Hamiltonian $H_{\rm eff}$ provide footprints of the $0$- and $\pi$-modes. However, one cannot obtain complete topological characterization by employing conventional static topological invariants using the Floquet operator or $H_{\rm eff}$. This can be understood from the fact that topological invariants of an energy band count the difference between the number of topologically protected modes below and above said band. As a consequence, the presence of an equal number of $0$- and $\pi$-modes cause static invariants to vanish, thereby designating the band as topologically trivial, even though the system harbors boundary states. To solve this issue, one can define a ``periodized'' evolution operator $U_\epsilon(t)$, which can be employed to topologically characterize the boundary states at the quasienergy gap $\epsilon$~\cite{Rudner2013,CarpentierPRL2015,Yan2017}. For a two-dimensional~(2D) system with broken time-reversal symmetry~(TRS), one can then compute a homotopy-based winding number by evaluating integrals over time and two other spatial variables~\cite{Rudner2013}. However, evaluating this winding number is numerically very challenging, particularly in the presence of disorder, or generally then translational symmetry is broken where instead a very large unit cell needs to be considered for the spatial variables~\cite{TitumPRX2016}. Moreover, the periodized evolution operator has to be constructed each time to compute topological invariants for different topological gaps $\epsilon$. Hence, seeking a formalism to construct topological invariants using the Floquet operator or $H_{\rm eff}$ directly, and thereby avoiding $U_\epsilon(t)$, would be very useful.

To this end, we utilize in this work the \textit{spectral localizer}~\cite{LoringAnnPhys2015,loring2017finitevolume, loring2019guide}, which is a real space tool that essentially measures whether states at a given energy are close to an atomic insulator without closing the bulk gap or breaking any symmetry. It has attracted increasing attention and has already been employed in a few systems where conventional topological invariants face challenges, such as gapless topological systems~\cite{CerjanPRB2022}, disordered semi-metals~\cite{HermannEPL2022,FrancaPRB2024}, photonic systems~\cite{CerjanNanoPhoton2022,DixonPRL2023}, and quasicrystals~\cite{FulgaPRL2016,LoringJMP2019}. For a static and translationally invariant system, we can compute well-established topological invariants for all the ten symmetry classes of the Altland–Zirnbauer~(AZ) classification~\cite{AltlandPRB1997,LoringAnnPhys2015}. In these established cases, the spectral localizer offers a complementary tool, as it is defined in real space and is also energy-resolved. It is this energy-resolved nature that also makes the spectral localizer a very promising tool for constructing topological invariant in driven systems to easily access also their anomalous $\pi$-modes.

Very recently, the spectral localizer has been employed for topological states at different quasienergies in a quasiperiodically driven one-dimensional~(1D) Kitaev chain~\cite{qi2024realspace} and also to study anomalous $\pi$-modes in a driven Chern insulator~\cite{LiuPRB2023}. However, the former proposal is based on particle-hole symmetry, and the latter relies on chiral symmetry. Also, no attention has been given to the effect of disorder on the topological invariants, despite it being ever-present in physical systems. In this work, we, therefore, set out to answer the following pertinent questions: (a) Is it possible to compute a topological invariant for a driven system from the effective Floquet Hamiltonian $H_{\rm eff}$, without relying on any symmetry and using the spectral localizer? (b) Can we employ this topological invariant for a disordered system and even predict disorder-induced phase transitions? In this work, we answer both of these intriguing questions in an affirmative way and, importantly, also establish that the spectral localizer is a very efficient numerical tool for driven systems. Together, this demonstrate the spectral localizer as an invaluable tool for topological Floquet systems.

In particular, in this work, we consider a driven 2D Chern insulator in class A in the AZ classification, i.e.~with no symmetries, in the form of periodic steps. This driven system hosts both $0$- and $\pi$-modes, which has been numerically cumbersome to classify using existing methods~\cite{Rudner2013,TitumPRX2016}, as they involve expensive numerical computation of the periodized evolution operator, with integrals over both time and spatial variables. By employing the spectral localizer to extract an energy-resolved topological invariant in real space, we are able to topologically characterize the $0$ and the $\pi$ boundary modes individually, fully unambiguously, and in a numerically efficient way. We further establish that in the presence of an onsite random disorder potential, the system can transmute into a disorder-induced Floquet topological Anderson insulator~(FTAI)~\cite{TitumPRL2015,RoyDisorderPRB2016,TitumPRX2016,TitumPRB2017,ChenPRB2018,DuPRB2018,MenaPRB2019,NingHallPRB2022} with the same energy resolved topological invariant based on the spectral localizer capturing the FTAI phase. 
To further corroborate our findings, we additionally consider a topological 1D driven system in the chiral AIII class to showcase how to extract a topological invariant based on the spectral localizer also in a 1D system with minimal symmetries, including in the presence of disorder. Our results, therefore, establish topological invariants directly based on the spectral localizer as a numerically efficient way to characterize driven and also disordered systems.

The rest of the work is organized as follows. In Sec.~\ref{Sec:drivenchern}, we study a driven 2D Chern insulator. We first introduce the spectral localizer and then extract an associated energy-resolved topological invariant in Sec.~\ref{SubSec:SpectralLocalizer2D}. We introduce the driving protocol in Sec.~\ref{SubSec:DrivingProtocol2D}, followed by presenting and discussing the resulting phase diagram and topological characterization for both the clean and disordered system based on the spectral localizer topological invariant in Sec.~\ref{SubSec:phase2D}-\ref{SubSec:disorder2D}. In Sec.~\ref{Sec:drivenSSH}, we perform a similar study of the topological characterization of a clean and disordered driven 1D chiral symmetric system. Finally, we summarize our work and provide a brief outlook in Sec.~\ref{Sec:summary}.

\section{Driven 2D Chern insulator} \label{Sec:drivenchern}
In this section, we study a driven 2D Chern insulator in class A of the AZ classification to derive a topological invariant based on the spectral localizer for systems without any inherent symmetries. We start by discussing the spectral localizer and its properties, including the localizer gap, and then introduce the relevant topological invariant, an energy-resolved real-space Chern marker to be employed for the topological characterization. We then specify the specific, but still generic, 2D Chern insulator we study in this work, using periodic steps. We finally report our results, both in the clean and disordered cases, using both the spectral localizer gap and establish the topological invariant, to both unambiguously and efficiently characterize the system.

\subsection{Spectral localizer and topological invariant} \label{SubSec:SpectralLocalizer2D}
We start by discussing the general properties of the spectral localizer~\cite{LoringAnnPhys2015, loring2017finitevolume, loring2019guide}. The spectral localizer $L_{\vect{\lambda}}(X_j,H)$ is a composite Hermitian operator consisting of the system's Hamiltonian $H$ and the position operators $X_j$, with $\vect{\lambda}=(x_j,E)$ where $x_j$ and $E$ are positions in real space and energy, respectively. Here the subscript $j$ runs over the spatial dimensions of the system. The spectral localizer measures a system's topology in real space by taking into consideration whether the Hamiltonian $H$ and the position operators $X_j$ can continue to commute, as long as there is no breaking of any additional symmetries or closing of the bulk gap~\cite{LoringAnnPhys2015,loring2017finitevolume,loring2019guide}. To this end, we introduce for a 2D system the spectral localizer $L_{x,y,E}(X,Y,H)$ in the Clifford representation as~\cite{LoringAnnPhys2015,loring2017finitevolume,loring2019guide} 
\begin{align}
    L_{x,y, E} \! \left(X,Y, H\right)\!=\!\kappa \left[ (X\! -\!x I) \tau_x \!+\!  (Y\! -\!y I)  \tau_y\right] \!+ \!(H \!-\! EI)  \tau_z.
    \label{localizer2D}
\end{align}
 Here, $X~(Y)$ and $x~(y)$ represent the position operator and position in real space in the $x~(y)$-direction, respectively, while the Pauli matrices $\vect{\tau}$ span the Clifford representation. The scaling constant $\kappa$ ensures compatible weight of the Hamiltonian $H$ and the position operators $X,Y$ and should be chosen within a permitted range~\cite{loring2017finitevolume, loring2020spectral}. In Appendix~\ref{App:A}, we study the topological invariant with variation of the scaling constant $\kappa$ to ensure a proper choice. Notably, we can compute the spectral localizer at any position $(x,y)$ in space and at any given energy $E$, which can even be chosen to be outside of the system's physical dimensions (such as in the surrounding vacuum) or spectrum. In particular, being able to compute the spectral localizer at any energy motivates us in this work to employ it for a driven system that can host topological boundary modes at different energies. 

An important quantity that we can directly extract from the spectral localizer is the localizer gap $\sigma_{\rm L}^{\vect{\lambda}}\left(X_j,H\right)$, defined as~\cite{CerjanJMP2023,WongPRB2023}
\begin{align}
    \sigma_{\rm L}^{\vect{\lambda}}\left(X_j,H\right)= {\rm min} \left( \lvert \sigma \left[L_{\vect{\lambda}}(X_j,H) \right] \rvert \right),
    \label{eq:Lgap}
\end{align}
where $\sigma \left[L_{\vect{\lambda}}(X_j,H) \right]$ represents the spectrum of $L_{\vect{\lambda}}(X_j,H)$. The localizer gap $\sigma_{\rm L}^{\vect{\lambda}}\left(X_j,H\right)$ provides us with information about the existence and location of topological boundary states at a given energy. If there exists a topological boundary state at $\vect{\lambda}_B$, the localizer gap $\sigma_{\rm L}^{\vect{\lambda}_B}\left(X_j,H\right)$ vanishes, while it remains finite for all other $\vect{\lambda}$. Thus, $\sigma_{\rm L}^{\vect{\lambda}_B}\left(X_j,H\right)$ can act as a ``local topological bandgap'' of the system, which only closes at the boundaries of the system if it hosts a topological boundary state, but remains finite elsewhere.

The next step is to also extract a topological index from the spectral localizer. Here we focus on a 2D system in class A of the AZ symmetry classification~\cite{AltlandPRB1997}, also known as the Chern class. For this system, the boundary modes are protected by only a bulk gap, as the system can break all symmetries. Such systems are topologically characterized by the Chern number, which is conventionally computed from the eigenstates of the system~\cite{fukui2005chern,loring2019guide}. The spectral localizer offers an alternate route without computing the system's eigenstates. In particular, it has been shown that a topological index, equivalent to the Chern number, can be computed from $L_{x,y, E} \left(X,Y, H\right)$ as~\cite{LoringAnnPhys2015}
\begin{align}
  C_{E} = \frac{1}{2}{\rm sig}[L_{x,y, E} \left(X,Y, H\right)], 
     \label{localChern}
\end{align}
where ${\rm sig}$ represents the signature of the matrix, which counts the difference between the number of positive and negative eigenvalues of that matrix. Here $C_E$ refers to the Chern number of the boundary state that crosses through $E$. As a consequence, $C_E$ is energy dependent, as well dependent on spatial position, and thereby becomes an energy-resolved Chern marker at each site.

In this work, we exploit the ``energy-filtering'' in the topological index or marker $C_E$ for a driven system, which exhibits topological boundary states at different energies. In particular, we replace the Hamiltonian $H$ in Eq.~(\ref{localChern}) with the effective Floquet Hamiltonian $H_{\rm eff}$ and generally put $E$ inside the quasienergy gap we are studying. The energy-resolved Chern marker $C_E$ also exhibits an interesting spatial distribution that can be connected directly with the localizer gap in Eq.~\eqref{eq:Lgap}: Where the energy-resolved Chern marker makes a transition from a finite value to zero at the boundaries of the system, the localizer gap also vanishes. Conversely, for a trivial system, the localizer gap is finite everywhere, and thus, the Chern marker remains at zero. We demonstrate these behaviors later in Fig.~\ref{Fig:2DChernAnalysis}. 

The computation of the signature employing the effective Floquet Hamiltonian for a driven system is numerically much more efficient compared to the standard treatment where a homotopy-based winding number involving three integrals is used~\cite{Rudner2013,TitumPRX2016}. We can make the computation even more efficient by noting that we can write $L_{x,y, E} \left(X,Y, H\right)=P_{L} D P_{U}$, with $P_{L}~(P_U)$ being a block lower~(upper) triangular matrix and $D$ a block diagonal matrix~\cite{ChadhaPRB2024}, given by
\begin{align}
    P_L &= 
    \begin{pmatrix}
        \mathbb{I} & \vect{0} \\
        \left(\tilde{X}+ i \tilde{Y} \right) \tilde{H}^{-1} & \mathbb{I}
    \end{pmatrix} , \quad
    P_U= 
    \begin{pmatrix}
        \mathbb{I} & \tilde{H}^{-1} \left(\tilde{X}- i \tilde{Y} \right)  \\
        \vect{0} & \mathbb{I}
    \end{pmatrix} , \non \\
    D &= 
    \begin{pmatrix}
        \tilde{H}   & \vect{0} \\
       \vect{0} & -\tilde{H} - \left(\tilde{X}+ i \tilde{Y} \right) \tilde{H}^{-1} \left(\tilde{X}- i \tilde{Y} \right) 
    \end{pmatrix} , 
\end{align}
where we have defined $\tilde{X}=(X\! -\!x I)$, $\tilde{Y}=(Y\! -\!y I)$, and $\tilde{H}=(H\! -\!E I)$. Further, $P_U=P_L^\dagger$ and thus this is equivalent to LDLT factorization, with $L_{x,y, E} \left(X,Y, H\right)=P D P^\dagger$, with $P=P_L$. Therefore, we can use Sylvester’s law of inertia to obtain ${\rm sig}[L_{x,y, E} \left(X,Y, H\right)]={\rm sig}[D]$~\cite{BuenoLAA2007}. This means the Chern number $C_E$ can be calculated as
\begin{align}
    C_E=\frac{1}{2} \left( {\rm sig} \left[H \right] -{\rm sig} \left[ \tilde{H} + \left(\tilde{X}+ i \tilde{Y} \right) \tilde{H}^{-1} \left(\tilde{X}- i \tilde{Y} \right) \right] \right) \ .
    \label{chernNoLDLT}
\end{align}
Here, Eq.~\eqref{chernNoLDLT} involves diagonalizing only a lower dimensional matrix, i.e.~the blocks of $D$, hence making the computation faster. With a fast and efficient way to calculate both a space- and energy-resolved Chern marker, we now have a tool to study Floquet systems in real space. To access pure bulk properties, we simply compute $C_E$ at each lattice site of a small box well inside the system and then take an average over the total number of sites entrapped inside said box.

\subsection{Driving protocol and effective Floquet Hamiltonian} \label{SubSec:DrivingProtocol2D}
Having introduced the spectral localizer and a topological invariant to study 2D Chern insulators, we introduce the particular system we are studying here. We use a driving protocol in the form of periodic steps as~\cite{ghosh2021systematic,ghosh2023HOTAI}
\begin{align}
    H(t)=& J_1 h_1^{\rm 2D} \  ,  \hspace{1cm} t \in \left[0,T/4\right]  , \nonumber \\
    =&J_2 h_2^{\rm 2D} \ , \hspace{1cm} t \in \left(T/4,3T/4 \right] ,  \nonumber \\
    =& J_1 h_1^{\rm 2D} \  , \hspace{1cm}   t \in \left(3T/4,T\right] ,
    \label{drive1}
\end{align}
over the period $T=2\pi/\Omega$, where the step Hamiltonians $h_{1,2}^{\rm 2D}$ in real space reads as
\begin{align}
    h_1^{\rm 2D}=& \sum_{\alpha,\beta} c_{\alpha,\beta}^\dagger \sigma_z c_{\alpha,\beta} \ , \non  \\
    h_2^{\rm 2D}=&  \frac{1}{2} \! \sum_{\alpha,\beta} c_{\alpha,\beta}^\dagger  \big( \sigma_z  c_{\alpha+1,\beta} \!+\! \sigma_z  c_{\alpha,\beta+1}  \!-\! i \sigma_x c_{\alpha+1,\beta} \!- \!i \sigma_y c_{\alpha,\beta+1}\big) \! \non \\
    &+ \!  {\rm H.c.}.
    \label{stepHam}
\end{align}
Here $c_{\alpha,\beta}$ creates an electron at a lattice site $(\alpha,\beta)$, with the $\alpha$ and $\beta$ indices representing $x$- and $y$-directions, respectively, while the Pauli matrices $\vect{\sigma}$ act on an orbital degrees of freedom. Here, $h_1^{\rm 2D}$ consists of only onsite terms and represents a trivial atomic insulator, while $h_2^{\rm 2D}$ includes only hopping terms (both real and imaginary) in both the $x$- and $y$-directions and manifests a Chern insulator phase with gapless edge states~\cite{ZhangSpincurrent2006}. Moreover, the tuning parameters $J_1$ and $J_2$ have the dimensions of energy, and we therefore use the dimensionless parameters $\left(J_1',J_2'\right)=\left(J_1T,J_2T\right)$ to control the strength of the drive. For simplicity, we use $T=2$ throughout this work. We here choose to work with the three-step drive in Eq.~\eqref{drive1} to have a symmetric driving protocol. However, we could alternatively employ a two-step drive, which would give a similar phase diagram~\cite{ghosh2021systematic}.

Considering the periodic step Hamiltonians in Eq.~\eqref{stepHam}, we construct the Floquet operator $U(T,0)$ as~\cite{GhoshJPCMReview2024}
\begin{align}
    U(T,0)={\rm TO} \exp \left[ -i \int_0^T  H(t') dt' \right] \ ,
    \label{FO}
\end{align}
where ${\rm TO}$ represents time-ordering. Using Eq.~\eqref{stepHam} we thus we obtain the Floquet operator as $U(T,0)= \exp \left( -i J_1 h_1^{\rm 2D} \frac{T}{4} \right) \exp \left( -i J_2 h_2^{\rm 2D} \frac{T}{2} \right) \exp \left( -i J_1 h_1^{\rm 2D} \frac{T}{4} \right)$.
We further consider the eigenvalue equation $U(T,0)\ket{\Psi_m}=\exp(-i E_m' T) \ket{\Psi_m}$, with $E_m'$ being the quasienergy of the state $\ket{\Psi_m}$ and $m$ the index of the state.  As we construct the Floquet operator in the time domain, the spectrum $E_m'$ is restricted to $[-\Omega/2,\Omega/2]$. However, in this work, we define the quasienergy as $E_m=E_m'T$, such that $E_m \in[-\pi,\pi]$. We can also straightforwardly obtain the effective Floquet Hamiltonian by numerically taking a logarithm of the Floquet operator matrix as $H_{\rm eff}=-i \ln \left[ U(T,0) \right]$. We then use the Floquet operator and $H_{\rm eff}$ to obtain the phase diagram of this system.

Owing to the simple form of the driving protocol, in this case, we can even analytically obtain the phase diagram in the $J_1 \mhyphen J_2$ plane based on earlier work~\cite{ghosh2021systematic,GhoshDynamical2022}. In particular, by considering the Bloch form of \Eq{stepHam}, i.e.~the momentum space Hamiltonians, and the corresponding Floquet operator, we obtain the condition for gap-closing around quasienergy $E=0$ and $E=\pi$ as $\lvert J_2' \rvert = \lvert J_1' \rvert/2 + n \pi $; with $n \in \mathbb{Z}$~\cite{ghosh2021systematic,GhoshDynamical2022}. These gap-closing conditions define the topological phase boundaries of the system. We plot these gap-closing lines (red and orange dashed lines) in Fig.~\ref{Fig:2DChern}.

\subsection{Results: Phase diagram} \label{SubSec:phase2D}
\begin{figure}
    \centering
    \subfigure{\includegraphics[width=0.39\textwidth]{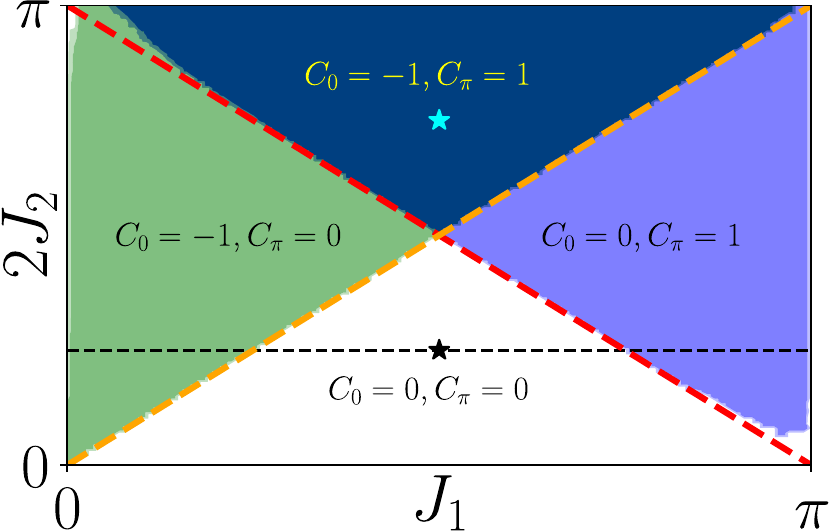}}
    \caption{Phase diagram of driven 2D Chern insulator as a function of $J_1$ and $J_2$ in terms of the energy-resolved Chern numbers $(C_0,C_\pi)$. Red and orange dashed lines represent the analytical bulk gap closing condition at quasienergy $0$ and $\pi$, respectively. Colored stars mark parameters used in Fig.~\ref{Fig:2DChernAnalysis}. Thin black line marks parameters used in Fig.~\ref{Fig:2DChernPhaseDisorder}. We use $\kappa=0.01$ and a system consisting of $24 \times 24$ lattice sites. Chern numbers $C_E$ are extracted as bulk values by averaging over a box of size $6 \times 6$ sites in the middle of the system.
   }
    \label{Fig:2DChern}
\end{figure}

In the rest of the section, we present our results for driven 2D Chern insulators. We start in this subsection by computing the energy-resolved Chern marker $C_E$ using the spectral localizer $L_{x,y, E} \left(X,Y, H\right)$ in Eq.~(\ref{localChern}) for the effective Floquet Hamiltonian. By calculating the marker for sites deep inside the bulk and then averaging over such bulk sites, the site-resolved marker becomes a single Chern number, still energy resolved. In particular, we focus on two energies in the known topological gaps, $E=0$ and $E = \pi$. We plot the result in Fig.~\ref{Fig:2DChern} in $J_1 \mhyphen J_2$-plane along with the earlier extracted analytical results. We find a phase diagram is divided into four parts with $(C_0,C_\pi)=(-1,0)$ (green), $(-1,1)$ (dark blue), $(0,1)$ (light blue) and $(0,0)$ (white). Here, finite $C_0$ or $C_\pi$ indicate the existence of non-trivial boundary modes at $E=0$ or $E=\pi$, respectively. We find that these numerically obtained topological phases match very well with the analytically obtained gap-closing lines (orange and red dashed lines). We only notice small discrepancies around the phase transition line, which we can attribute to finite-size effects and a tiny bulk gap. We are thus able to fully reproduce the phase diagram using numerically efficient energy-resolved Chern numbers, instead of using a cumbersome homotopy-based invariant~\cite{Rudner2013,TitumPRX2016}, thereby allowing analysis of a driven system to be more accessible.

\subsection{Results: Localizer gap and local Chern marker} 
\label{SubSec:gapmarker2D}
\begin{figure}
    \centering
    \subfigure{\includegraphics[width=0.49\textwidth]{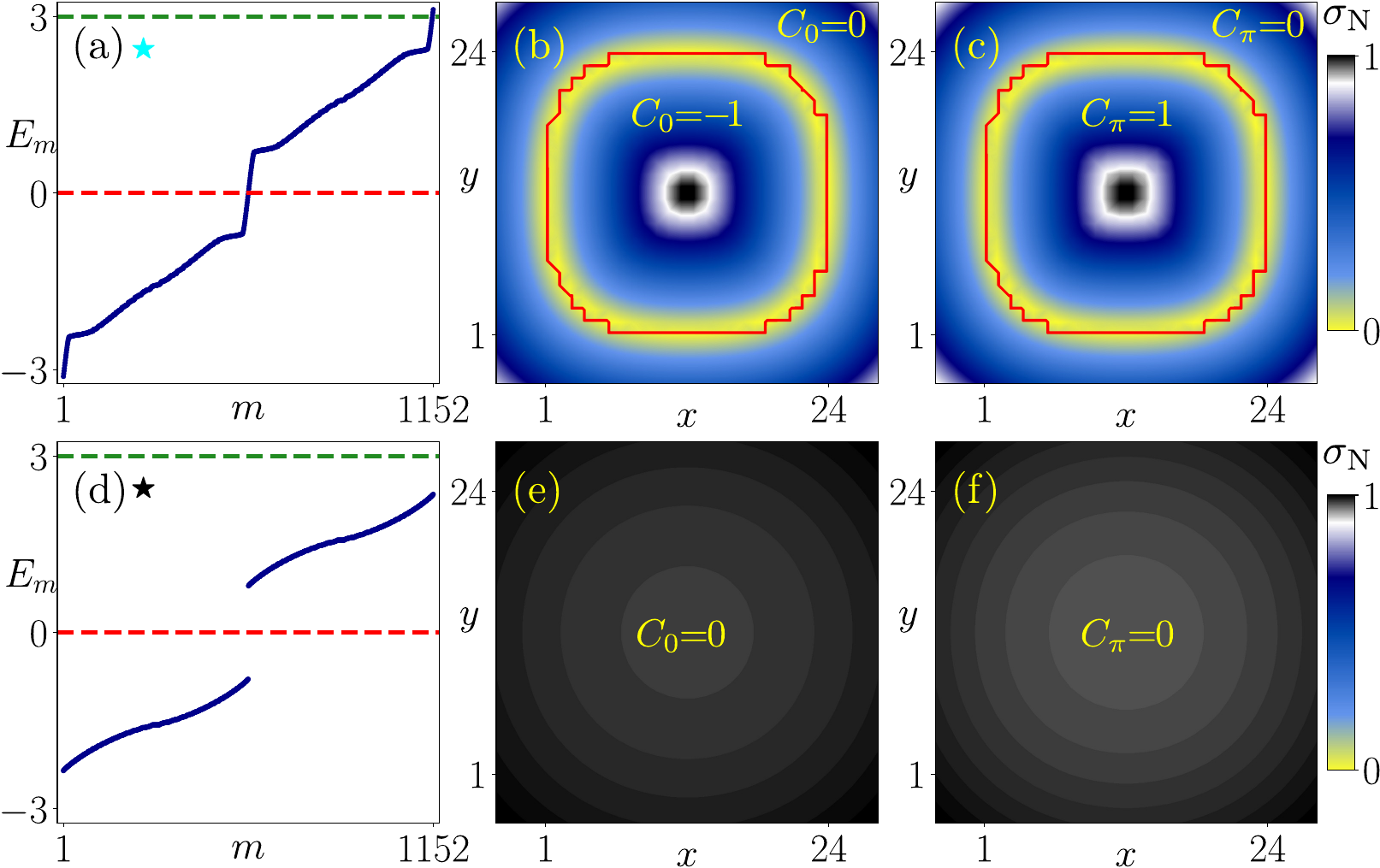}}
    \caption{(a) Quasienergy $E_m$ as a function of  state index $m$ for a non-trivial phase marked by cyan star, $(J_1,2J_2)=(\pi/2,\pi/4)$, in Fig.~\ref{Fig:2DChern}. Red and green dashed lines indicate quasienergies inside the $0$- and $\pi$-gaps, employed to compute the localizer gap and Chern marker. (b,c) Space-resolved normalized localizer gap $\sigma_{\rm N}^{x,y,E}\left(X,Y,H\right)$ throughout and beyond the system, computed inside the quasienergy gap at $E=0$ (b) and $E=\pi$ (c), with colorbar indicating the amplitude of $\sigma_{\rm N}$.
    Solid red lines separate regions with trivial and non-trivial Chern markers as indicated by labels $C_{0,\pi}$.
    (d-f) Same as (a-c) but for trivial phase marked by black star, $(J_1,2J_2)=(\pi/2,3\pi/4)$, in Fig.~\ref{Fig:2DChern}.
    We use $\kappa=0.01$ and a system consisting of $24 \times 24$ lattice sites (situated as $1 \leq x,y \leq 24$).
    }
    \label{Fig:2DChernAnalysis}
\end{figure}

Having demonstrated that the energy- and space-resolved Chern marker fully reproduces the phase diagram, we next analyze its connection to the localizer gap. For this purpose, we choose a point from the phase diagram in Fig.~\ref{Fig:2DChern} (cyan star), where the driven system hosts both $0$- and $\pi$-modes. We depict the corresponding quasienergy spectrum, i.e.~$E_m$ as a function of the state index $m$, of the effective Floquet Hamiltonian $H_{\rm eff}$ for a system obeying open boundary conditions~(OBC) in both directions in Fig.~\ref{Fig:2DChernAnalysis}(a). We see the existence of states crossing both the $0$- and $\pm \pi$-gaps, which are the topological boundary states. We then choose two energies (red and green dashed lines) in the middle of these two gaps to compute the localizer gap and the energy-resolved Chern marker. We plot the normalized localizer gap $\sigma_{\rm N}^{x,y,E}\left(X,Y,H\right)=\sigma_{\rm L}^{x,y,E}\left(X,Y,H\right)/{\rm max}\left[\sigma_{\rm L}^{x,y,E}\left(X,Y,H\right)\right]$ inside these $0$- and $\pi$-gaps in Figs.~\ref{Fig:2DChernAnalysis}(b) and (c), respectively, with the colorbar indicating the weight of $\sigma_{\rm N}^{x,y,E}\left(X,Y,H\right)$. In both plots, we observe that $\sigma_{\rm N}^{x,y,E}\left(X,Y,H\right)$ goes to zero sharply around the edges of the system (yellow stripe), while it remains finite inside the system as well as outside of the system. The closing of the localizer gap around the boundary provides us with an indication of the boundary states residing around that region. We also notice how the localizer gap is rounding off the corners of the system, just as a boundary state likely behave. The next step is associating a topological indicator with this localizer gap closing. To this end, we compute the energy-resolved Chern marker $C_E$ employing Eq.~(\ref{localChern}) on each site viewed in Figs.~\ref{Fig:2DChernAnalysis}(b,c). Note that this involves both sites within the system (here defined as sites $1 \leq x,y \leq 24$) and outside in the surrounding vacuum. We find a non-zero $C_0=-1$ in (b) and $C_\pi = 1$ in (c) in the system stretching to the red solid line, which marks the boundary to the $C_{0,\pm \pi} =0$ region. As seen, we find an essentially perfect agreement between the closing of the localizer gap (yellow stripe) and the spatially resolved topological invariant $C_E$ (red line).

Finally, we also consider the trivial phase. We demonstrate the quasienergy spectrum $E_m$ as a function of the state index $m$ of the effective Floquet Hamiltonian $H_{\rm eff}$ obeying OBC in Fig.~\ref{Fig:2DChernAnalysis}(d), corresponding to the black star in Fig.~\ref{Fig:2DChern}. It is clear from the plot that the system does not possess any states in either the $0$- or $\pi$-gap. Moreover, the normalized localizer gap $\sigma_{\rm N}^{x,y,E}\left(X,Y,H\right)$ remains finite everywhere for both $0$- and $\pi$-gap, as seen from Figs.~\ref{Fig:2DChernAnalysis}(e) and (f), respectively. At the same time, both $C_0$ and $C_\pi$ are zero everywhere in Figs.~\ref{Fig:2DChernAnalysis}(e,f). Thus, we conclude from Fig.~\ref{Fig:2DChernAnalysis} that the spectral localizer captures both the existence and spatial position of boundary states through its localizer gap, defined in Eq.~\eqref{eq:Lgap}, and instills a topological invariant through the energy-resolved Chern marker $C_E$, defined in Eq.~\eqref{localChern}, also with full spatial resolution. Additionally, and important for driven systems, $C_E$ per definition assigns an invariant to each quasienergy gap, also those at finite energy.

\subsection{Results: Effects of disorder}
\label{SubSec:disorder2D}
So far, we have only considered a driven system without disorder but with a topological invariant specified in real space, also disordered systems are within reach. The presence of disorder allows us to investigate the stability of the energy-resolved Chern numbers. Moreover, it is known that disorder can also induce non-trivial topology in an otherwise trivial system. Such disorder-induced driven topological phases are called Floquet topological Anderson insulator~(FTAI) and have previously been reported ~\cite{TitumPRL2015,RoyDisorderPRB2016,TitumPRX2016,TitumPRB2017}. Thus, it would be intriguing to investigate if the spectral localizer-based Chern numbers can also predict the FTAIs. To this end, we consider the effect of disorder on the Chern marker $C_E$ and the resulting Chern number within individual disordered samples and establish that it still manages to also capture disordered topological phases. On this account, we incorporate onsite random disorder by modifying the step Hamiltonian $h_1^{\rm 2D}$ as
\begin{align}
    \tilde{h}_1^{\rm 2D}&= \sum_{\alpha,\beta} c_{\alpha,\beta}^\dagger \left[1+\frac{V_{\alpha \beta}}{J'_1} \right] \sigma_z c_{\alpha,\beta} \ , 
    \label{disorderstepHam}
\end{align}
where $V_{\alpha \beta}$ is uniformly distributed in $\left[-w/2,w/2 \right]$, with $w$ being the strength of the disorder. 
To study disorder, we both study individual disorder configurations, equivalent to individual samples, and disorder average over $50$ random configurations, which produces reasonably well-converged results for the average behavior. To access bulk properties, we here consider a system of size $20 \times 20$ site, but only access the energy-resolved Chern marker by averaging it over a box of size $6 \times 6$ lattice sites in the middle of the system for each random configuration. 

\begin{figure}
    \centering
    \subfigure{\includegraphics[width=0.48\textwidth]{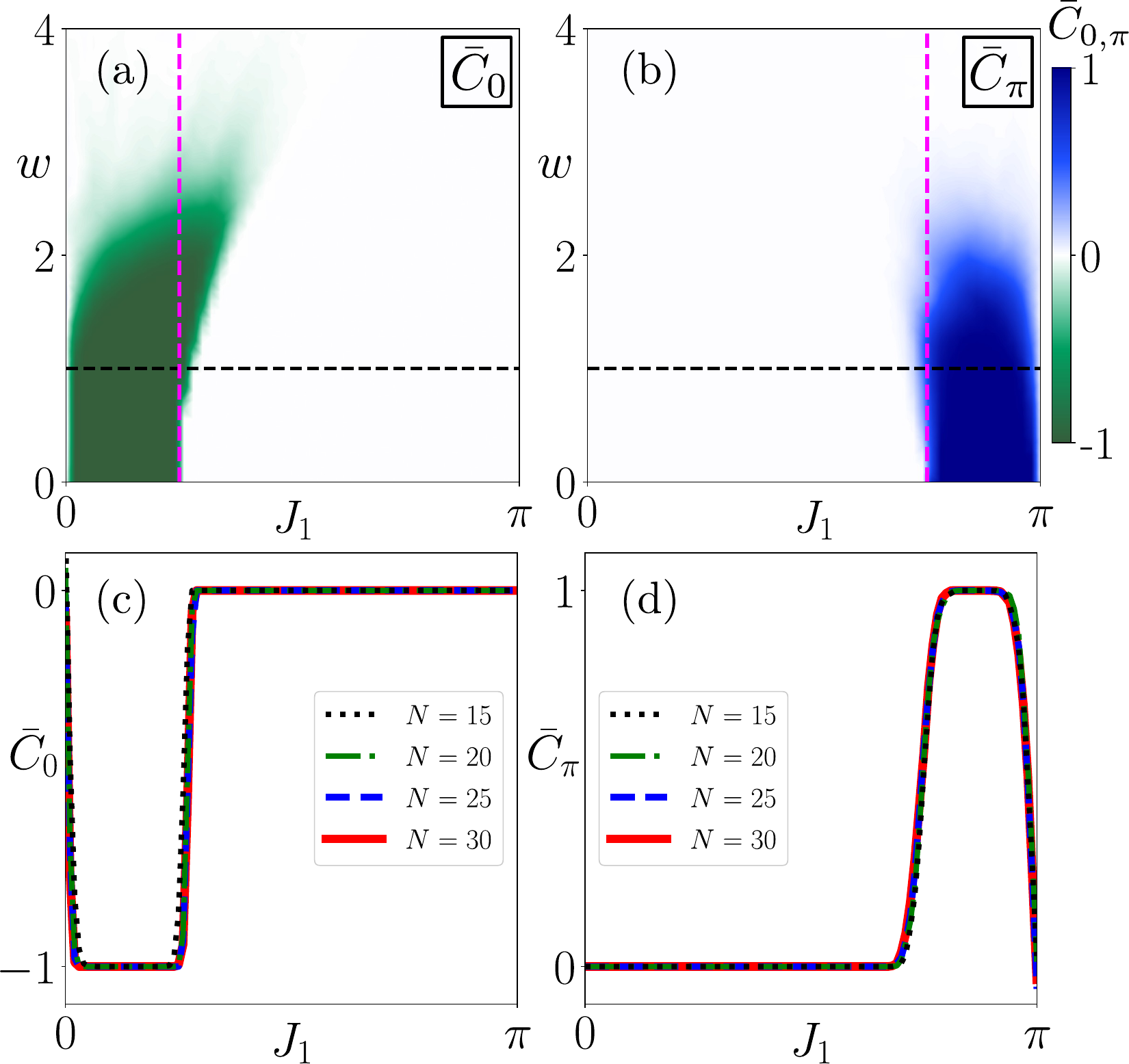}}
    \caption{Disorder-averaged $\bar{C}_0$ (a) and $\bar{C}_\pi$ (b), in the $J_1 \mhyphen w$-plane for $J_2=\pi/8$ ($J_1,J_2$ values marked by dashed black line in Fig.~\ref{Fig:2DChern}).  Magenta dashed lines denote the phase transition point for a clean driven system. Disorder-averaged $\bar{C}_0$ (c) and $\bar{C}_\pi$ (d) for different system sizes ($N \times N$) as a function of $J_1$ for a fixed $w$ (marked by dashed black lines in (a,b)). We use $\kappa=0.01$, 50 disorder configurations, and a system consisting of $20 \times 20$ lattice sites for (a,b). Chern numbers $\bar{C}_E$ are extracted as bulk values by averaging over a box of size $6 \times 6$ sites in the middle of the system for each disorder configuration.
    }
    \label{Fig:2DChernPhaseDisorder}
\end{figure}

To elucidate the disorder behavior we consider a fixed value of $J_2$, indicated by black dashed line in Fig.~\ref{Fig:2DChern}, and compute the disorder-averaged Chern numbers $\bar{C}_E$ at $E = 0,\pi$ in the $J_1 \mhyphen w$ plane in Fig.~\ref{Fig:2DChernPhaseDisorder}. For zero disorder ($w=0$), we find both finite $C_0=-1$ (green) and $C_\pi = 1$ (blue) phases, as well as a trivial phase (white), with the transition marked (dashed magenta lines).
The overall observation of the behavior with finite disorder is two-fold: we find a disorder-driven topological to non-topological phase transition and also a disorder-induced topological phase transition. 
To be precise, in Fig.~\ref{Fig:2DChernPhaseDisorder}(a), we observe that the system transmutes to a trivial topological phase with $\bar{C}_0$ with the increase of the disorder potential strength $w$. We note that a lighter green color means that some disorder configurations (i.e.~samples) have $C_0 = -1$ and are thus topological, while other configurations have $C_0 = 0$ and are trivial, thus resulting in non-quantization of $\bar{C}_0$. We see a similar disorder-driven topological to trivial phase transition also in Fig.~\ref{Fig:2DChernPhaseDisorder}(b) for the anomalous Floquet state ($\pi$-modes), roughly setting in at the same disorder strength.
Fascinatingly, disorder can also induce a topological phase transition. In the driven system, the topological phase set in at the magenta dashed lines for zero disorder in Fig.~\ref{Fig:2DChernPhaseDisorder}. Still, there exist solid green and blue regions beyond these lines. In these region we have a disorder-driven FTAI \cite{TitumPRL2015,RoyDisorderPRB2016,TitumPRX2016,TitumPRB2017}. To emphasize, the system is topologically trivial in the absence of disorder but exhibits a non-trivial topological phase when both periodic drive and random disorder are present. We note that this behavior is most visible for finite $\bar{C}_0$ in \ref{Fig:2DChernPhaseDisorder}(a) but also present for finite $\bar{C}_\pi$ in \ref{Fig:2DChernPhaseDisorder}(b). 

Finally, to understand if there exists any system size dependence on the disorder-averaged topological invariants, we also compute $\bar{C}_{0,\pi}$ for different system sizes $N$, such that the system consists of $N \times N$ lattice sites. In Figs.~\ref{Fig:2DChernPhaseDisorder}(c) and (d) we plot $\bar{C}_{0}$ and $\bar{C}_{\pi}$, respectively, as a function of $J_1$ for several different $N$ at a fixed value of $w=1$ (depicted by black dashed lines in Figs.~\ref{Fig:2DChernPhaseDisorder}(a,b)). We do not observe any drastic behavior for different system sizes. In fact, except for $N=15$, the Chern numbers $\bar{C}_{0,\pi}$ for all the other systems sizes fall on each other. Thus, we conclude that our results are not an artifact of a finite-size effects.

Our disorder results show that the spectral localizer-based topological invariant is not only capable of topologically characterizing the $0$- and $\pi$-modes distinctly in clean driven systems, but it also performs well in the presence of disorder and can intriguingly even capture the disorder-induced FTAI. 
Importantly, the calculation of $\bar{C}_E$ here only relies on the eigenvalues of the spectral localizer matrix and is thus very efficient even in the presence of disorder. This should be compared to the traditional homotopy-based winding number \cite{TitumPRX2016}, which involves a much more numerically expensive computation of eigenstates and integrals, especially for disordered systems or any other system where translation symmetry is broken.

\section{Driven 1D chiral symmetric system} \label{Sec:drivenSSH}
Having investigated a driven Chern insulator, we in this section supplement our results in two dimensions with complementary results for a 1D system. In one dimension, the static Hamiltonian, however, needs to belong to a different symmetry class than our treatment in two dimensions in the previous section, as the A class is not topological in one dimension. Here, we consider a driven chiral symmetric 1D system within class AIII in the AZ classification, as it only preserves the chiral symmetry. We note that the particle-hole symmetry case has also been treated very recently~\cite{qi2024realspace}. The chiral symmetry makes any such system also naturally host an orbital, often called sublattice, degree of freedom, just as in the previous section for the 2D Chern insulator.
As a consequence, the topological invariant is necessarily also different. In this section, we start by first presenting the topological invariant that is possible to extract from the spectral localizer for a driven chiral symmetric system in one dimension. We then again present the specific driving protocol and step Hamiltonians before presenting our results in both the clean and disordered limits.

\subsection{Topological invariant}
The spectral localizer for a 1D system reads as~\cite{LoringAnnPhys2015,loring2017finitevolume,loring2019guide}
\begin{align}
    L_{x, E}\left(X, H\right)= \kappa  (X -x I) \tau_x + (H - EI)  \tau_y .
    \label{localizer1D}
\end{align}
Similar to two dimensions, we can also define the localizer gap as $\sigma_{\rm L}^{x,E}\left(X,H\right)= {\rm min} \left( \lvert \sigma \left[L_{x,E}(X,H) \right] \rvert \right)$~\cite{LoringAnnPhys2015,loring2017finitevolume,loring2019guide}. The localizer gap vanishes at the ends of the system, i.e.~at the zero-dimensional~(0D) endpoints, if there exists a topological boundary state. 
Dealing now with a system with chiral symmetry, the topological index has to be defined differently from the 2D case in Sec.~\ref{Sec:drivenchern}. For a static 1D chiral symmetry a topological index $\nu$ can instead be computed as~\cite{LoringAnnPhys2015}
\begin{align}
     \nu={\rm sig}[(X+i H)S]  . 
     \label{localwind1D} 
\end{align}
Here, $S$ represents the chiral symmetry operator, such that $SHS=-H$, and thus the matrix $(X+i H)S$ is Hermitian. However, unlike the energy-resolved Chern marker in Eq.~(\ref{localChern}), there is no energy-filtering in the definition of $\nu$. Therefore, we cannot directly employ Eq.~(\ref{localizer1D}) to compute the topological invariant for a driven system that hosts both $0$- and $\pi$-modes. However, the chiral symmetry has previously been utilized to compute the winding number ~\cite{AsbothPRB2014}. Here, we use the same idea to compute the topological index, but now employed to the spectral localizer.

The chiral symmetry allows us to define two operators: $U_a=S U_c^\dagger S U_c$ and $U_b=U_c S U_c^\dagger S$, with the time-evolution operator $U_c=U(T/2,0)$. Both operators $U_{a,b}$ transforms under chiral symmetry as $SU_{a,b}S=U_{a,b}^\dagger$~\cite{AsbothPRB2014,ghosh2023HOTAI}, which is the same for the Floquet operator $U(T,0)$ transforming under chiral symmetry. Thus, we can define two effective Hamiltonians corresponding to these operators: $H_{\rm eff}^{a,b}=- i \ln U_{a,b}$ such that these Hamiltonians transforms under chiral symmetry as $S H_{\rm eff}^{a,b} S=-H_{\rm eff}^{a,b}$. Based on this we employ Eq.~(\ref{localwind1D}) to compute two winding numbers associated with $H_{\rm eff}^{a,b}$, respectively, as
$\nu_{a,b}={\rm sig}[(X+i H_{\rm eff}^{a,b})S]$. It has been shown that $\nu_{a,b}$ provides the difference between the total number of modes at each boundary per sublattice~\cite{AsbothPRB2014,ghosh2023HOTAI}. As a consequence, $\nu_{a,b}$ individually do not provide information about the number of $0$- and $\pi$-modes. 
Still, by studying the location of the  $0$- and $\pi$-modes in sublattice space, it has been shown that a combination of $\nu_{a,b}$ provides the number of $0$- and $\pi$-modes, given as~\cite{AsbothPRB2014,ghosh2023HOTAI}
\begin{align}
    W_0 = \frac{\nu_a+ \nu_b}{2}, \  {\rm and} \ W_\pi = \frac{\nu_a- \nu_b}{2}\ .
    \label{localwind1DFloquet}
\end{align}
Here, $W_0~(W_\pi)$ counts the number of $0~(\pi)$-modes per edge and thus can be employed to topologically characterize a driven system that hosts both $0$- and $\pi$-modes due to the native energy-resolution of the spectral localizer. Furthermore, $W_{0,\pi}$ has full spatial resolution, which, if utilized, makes it a real-space local winding marker for a driven 1D chiral system, which, when averaged over the bulk, using a small box well inside the system still keeping OBC, generate the bulk winding number, resolved individually for the $0$- and $\pi$-modes. In particular, for a topologically non-trivial phase, both $W_{0,\pi}$ are only finite inside the chain and jump to $0$ outside the chain. In the following, we utilize the $W_{0,\pi}$ winding marker and its average bulk value as the winding number to study a driven 1D chiral system and clearly distinguish between the $0$- and $\pi$-modes.

\subsection{Driving protocol and step Hamiltonians}
For simplicity, we study as a driven 1D chiral system the same driving protocol as introduced in Eq.~(\ref{drive1}) for the 2D system, but now use the step Hamiltonians
\begin{align}
    h_1^{\rm 1D}&= \sum_{\alpha} c_{\alpha}^\dagger \sigma_x c_{\alpha} \ , \non  \\
    h_2^{\rm 1D}&=  \frac{1}{2}  \sum_{\alpha} c_{\alpha}^\dagger  \big( \sigma_x   - i \sigma_y \big) c_{\alpha+1} + {\rm h.c.},
    \label{stepHam1D}
\end{align}
where $c_\alpha$ is the electronic creation operator that creates an electron at a lattice site $\alpha$, where the index $\alpha$ runs over the length of the chain. Again, the Pauli matrices encode an orbital or sublattice degree of freedom.
Notably, both the step Hamiltonians $h_{1,2}^{\rm 1D}$ preserves chiral symmetry: $S h_{1,2}^{\rm 1D}S=-h_{1,2}^{\rm 1D}$ with $S={\rm diag}\left(1,-1, \cdots,1, -1 \right)$. The step Hamiltonian $h_1^{\rm 1D}$ represents a band insulator, while $h_2^{\rm 1D}$ resembles the seminal Su–Schrieffer–Heeger model, which hosts 0D end states. We then employ Eq.~(\ref{FO}) to compute the Floquet operator $U(T,0)$ and $U_c=U(T/2,0)$ using the step Hamilonians $h_{1,2}^{\rm 1D}$ in Eq.~(\ref{stepHam1D}) within the driving protocol \ref{drive1}. Similar to in two dimensions, we can analytically obtain the gap-closing condition around quasienergies $E=0,\pi$ as $\lvert J_2' \rvert = \lvert J_1' \rvert + n \pi$ with $n \in \mathbb{Z}$~\cite{ghosh2021systematic,GhoshDynamical2022}. Finally, we use Eqs.~(\ref{localizer1D}), (\ref{localwind1D}), and (\ref{localwind1DFloquet}) to obtain the winding markers $W_0$ and $W_\pi$.

\subsection{Results: Phase diagram, localizer gap, and local winding marker}
\begin{figure}
    \centering
    \subfigure{\includegraphics[width=0.49\textwidth]{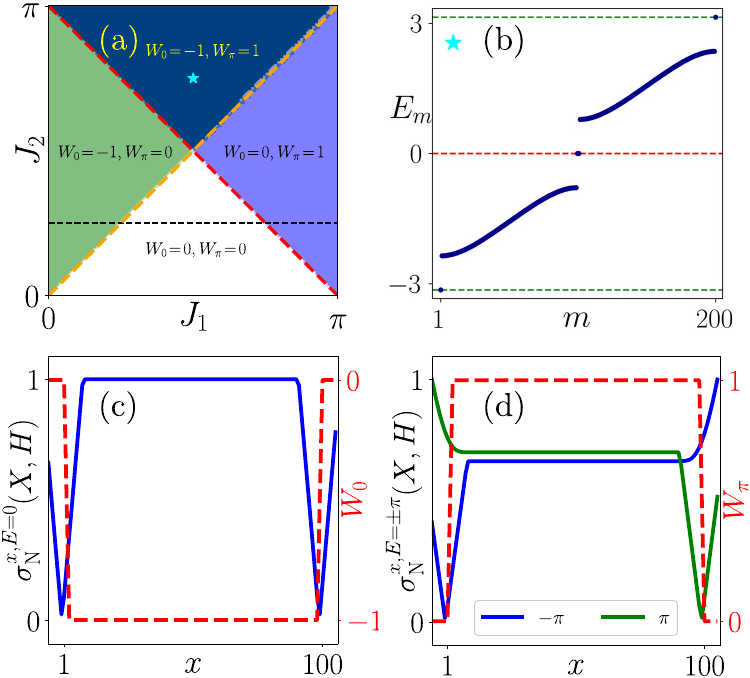}}
    \caption{(a) Phase diagram of driven 1D chiral symmetry system as a function of $J_1$ and $J_2$ plane in terms of energy-resolved winding numbers $(W_0,W_\pi)$. Red and orange dashed lines represent analytical bulk closing conditions at quasienergy $\pi$ and $0$, respectively. Cyan star and dashed black line mark parameters used in (b-d) and Fig.~\ref{Fig:1DSSHDisorder}, respectively.
    We use a chain consisting of $300$ lattice sites. Winding numbers $W_E$ are extracted as bulk values by averaging over a box of size $41$ sites in the middle of the system. (b) Quasienergy $E_m$ as a function of state index $m$ for a non-trivial phase marked by cyan star, $(J_1,2J_2)=(\pi/2,3\pi/4)$, in (a). (c,d) Space-resolved normalized localizer gap $\sigma_{\rm N}^{x,E}\left(X,H\right)$, computed at quasienergy $E=0$ (c,blue line) and $\pm \pi$ (d, blue and green lines), respectively. On the right axes are the topological winding markers $W_0$ (c, red dashed lines) and $W_\pi$ (d, red dashed lines). We use $\kappa=0.1$ and a chain of $100$ (situated as $ 1 \leq x \leq 100$) lattice sites (b-d).
    }
    \label{Fig:1DSSH}
\end{figure}

Having introduced the topological invariant, localizer gap, driving protocol, and step Hamiltonians above, we first calculate the phase diagram of this driven 1D system and present the results in Fig.~\ref{Fig:1DSSH}(a). 
We here investigate the winding numbers $W_0$ and $W_\pi$ in the parameter space spanned by the control parameters $J_1$ and $J_2$. The phase diagram resembles what we obtain in Fig.~\ref{Fig:2DChern} for the 2D driven system. It is divided into four different topological regimes: $(W_0,W_\pi)=(-1,0),~(-1,1),~(0,1),~{\rm and}~(0,0)$. The transitions between these numerically calculated phases coincide extremely well with the red and orange dashed lines, which represent the gap closing at quasienergies $\pi$ and $0$, respectively, which have already been obtained analytically~\cite{ghosh2021systematic,GhoshDynamical2022}. We attribute the very minor discrepancies to finite-size effects.

Next, we focus on a single parameter point marked by the cyan star in the phase diagram Fig.~\ref{Fig:1DSSH}(a) and utilize a chain of length $100$ sites. For this set of parameter values, we show the quasienergy spectrum $E_m$ as a function of the state index $m$ in Fig.~\ref{Fig:1DSSH}(b). We can clearly see the existence of modes at $E_m=0,\pm \pi$. We check that all these states are localized at the ends of the 1D chain. Here, we obtain two modes at quasienergy $E=0$ and one mode each at $E=-\pi$ and $\pi$, but we can bring back the mode at $-\pi$ to $\pi$ (or vice-versa) since they represent the same quasienergy in the FBZ. However, since the modes appearing at quasienergy $-\pi$ and $\pi$ are localized at two opposite ends of the chain, this distinction is manifested in the localizer gap. Similar to two dimensions, here also we employ the normalized localizer gap defined as $\sigma_{\rm N}^{x,E}\left(X,H\right)=\sigma_{\rm L}^{x,E}\left(X,H\right)/{\rm max}\left[\sigma_{\rm L}^{x,E}\left(X,H\right)\right]$. In Fig.~\ref{Fig:1DSSH}(c), we plot the normalized localizer gap $\sigma_{\rm N}^{x,E}\left(X,H\right)$ (blue line) for the states at quasienergy $E=0$ as a function of the real space direction $x$. As seen, $\sigma_{\rm N}^{x,E=0}\left(X,H\right)$ vanishes at both ends of the chain, exactly where the end states exist, but remains finite elsewhere. On the right axis of Fig.~\ref{Fig:1DSSH}(c), we additionally depict the winding marker $W_0$ (red dashed line). We find that $W_0$ stays at $-1$ inside the bulk of the chain, while it jumps to $0$ at the ends of the chain, exactly following the jump of the localizer gap.

Furthermore, in Fig.~\ref{Fig:1DSSH}(d), we show $\sigma_{\rm N}^{x,E}\left(X,H\right)$ for the states at quasienergy $E=-\pi$ (blue line) and $E=\pi$ (green line). We see that for the state at quasienergy $-\pi~(\pi)$, $\sigma_{\rm N}^{x,E}\left(X,H\right)$ vanishes only at the left~(right) end of the chain. This displays an important aspect of the spectral localizer, that it carries information regarding the localization of the boundary states, even those appearing at the same FBZ-based quasienergies $E= \pm \pi$. At the same time, we demonstrate the winding marker $W_\pi$ on the right axis of Fig.~\ref{Fig:1DSSH}(d) (red dashed line). It follows the localizer gap in the same fashion as for the zero-modes. In particular, $W_\pi$ is $1$ inside the bulk of the system and $0$ elsewhere. Taken together, results in Fig.~\ref{Fig:1DSSH} demonstrate that the spectral localizer can be employed successfully to topologically characterize a driven chiral symmetric system in one dimension. Both the localizer gap and the extracted winding markers are tools that fully capture both energy and spatial resolution of the boundary states, clearly resolved as $0$- and $\pi$-modes, and thus the topological region.

\subsection{Results: Effects of disorder}
Next, we consider the effect of onsite random disorder potential. Similar to the 2D disordered driven system, we here consider onsite disorder in the first step Hamiltonian as $\tilde{h}_1^{\rm 1D}= \sum_{\alpha} c_{\alpha}^\dagger \left[1+\frac{V_{\alpha}}{J'_1} \right] \sigma_z c_{\alpha}$; where again $V_\alpha$ is uniformly distributed in the range $\left[-w/2,w/2\right]$ and $w$ is the strength of the disorder. Note that the system maintains chiral symmetry even in the presence of disorder. We here consider a chain consisting of $100$ lattice sites but only access the energy-resolved winding markers for a smaller box of sites ($81$ sites) inside the bulk of the chain, which, when averaged, gives us a bulk winding number for that particular sample. We repeat this for $50$ different random disorder configurations to obtain the disorder-averaged winding numbers $\bar{W}_{0,\pi}$.

\begin{figure}
    \centering
    \subfigure{\includegraphics[width=0.48\textwidth]{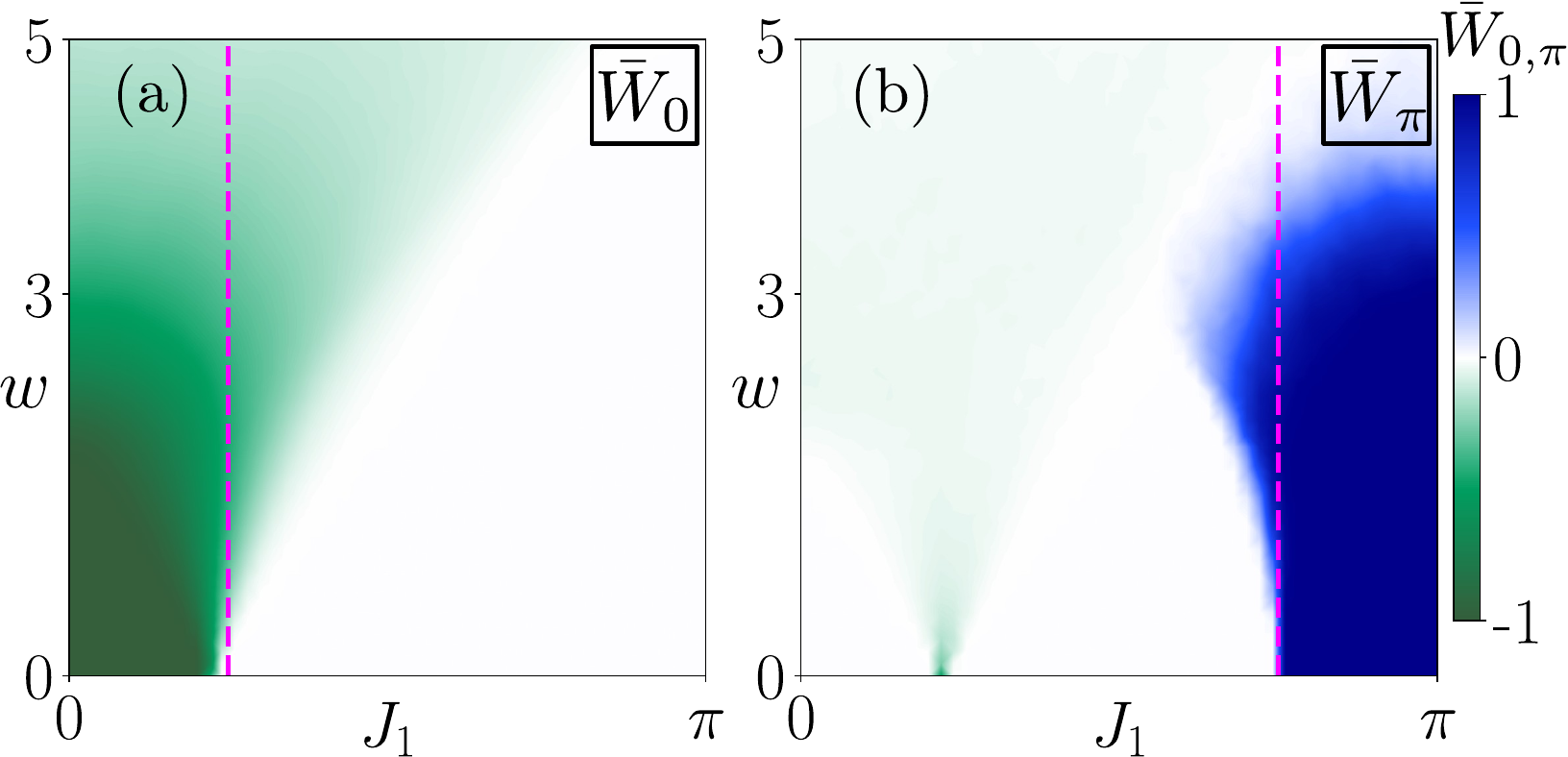}}
    \caption{Disorder-averaged winding numbers $\bar{W}_0$ (a) and $\bar{W}_\pi$ (b) in the $J_1 \mhyphen w$-plane for $J_2=\pi/4$ ($J_1,J_2$ values marked by dashed black line in Fig.~\ref{Fig:1DSSH}(a)). Magenta dashed lines denote the phase transition point for a clean driven system. We use $\kappa=0.1$, and we consider a chain consisting of $100$ lattice sites and $50$  disorder configurations. Winding numbers $\bar{W}_{0,\pi}$ are extracted as bulk values by averaging over a box size of $81$ sites in the middle of the chain for each disorder configuration.
    }
    \label{Fig:1DSSHDisorder}
\end{figure}

We plot the resulting disorder-averaged winding number $\bar{W}_0$ and $\bar{W}_\pi$ in the $J_1 \mhyphen w$ plane in Figs.~\ref{Fig:1DSSHDisorder}(a) and (b), respectively, for a fixed value of $J_2$, indicated by the dashed black in Fig.~\ref{Fig:1DSSH}(a). The colorbar denotes winding numbers such that $\bar{W}_0=-1$~(dark green) and $\bar{W}_\pi=1$~(dark blue), while the vertical magenta dashed line represents the phase transition line in the absence of disorder, i.e.~at $w=0$. From both Figs.~\ref{Fig:1DSSHDisorder}(a,b), we observe that the disorder-averaged winding numbers are still quantized for a small amplitude of the disorder potential, signifying that all samples are topological. With the increase of $w$, winding numbers tend to become less quantized and fade away, see light green and blue colors, indicating a disorder-driven topological to non-topological transition in more and more of the individual samples. Furthermore, we also observe the topological phase extending to the right and left of the vertical magenta dashed line in Figs.~\ref{Fig:1DSSHDisorder}(a) and (b), respectively, although less vivid in (a). This extended region indicates a disorder-induced topological phase, a signature of FTAI. 
As a side remark, we also observe in Fig.~\ref{Fig:1DSSHDisorder}(b) a light green shade around the point that coincides with the phase transition line marked for a clean driven system in Fig.~\ref{Fig:1DSSHDisorder}(a). We believe this discrepancy arises due to a numerical issue that occurs while computing $\nu_{a,b}$ in Eq.~(\ref{localwind1DFloquet}), owing to a very small bulk gap. Notably, this light green region in Fig.~\ref{Fig:1DSSHDisorder}(b) has no physical significance, but it is rather just a numerical artifact. 
Thus, the winding numbers $\bar{W}_{0,\pi}$ are capable of characterizing the 1D driven chiral symmetric system in the presence of disorder and can also capture disorder-induced topological phases, similar to the behavior for the driven 2D Chern insulator in Fig.~\ref{Fig:2DChernPhaseDisorder}.

\section{Summary and Outlook} \label{Sec:summary}
To summarize, in this work, we utilize the spectral localizer to extract a space and energy-resolved topological invariant to characterize driven systems. We consider both a 2D driven Chern insulator belonging to class A, i.e.~with no symmetries, and show that we can compute an energy-resolved Chern number, a really local marker, that is capable of unambiguously characterizing both the $0$- and $\pi$-modes. We additionally connect the energy-resolved Chern marker with the localizer gap and demonstrate that the Chern marker changes at the same location in real space only where the localizer gap closes. Being a real space tool, we also utilize these invariants to study a disordered driven system. The Chern number not only works for a disordered system, but it also captures a disorder-induced FTAI phase. Furthermore, we additionally consider a driven 1D chiral symmetric system in class AIII. Here, we compute a spatially resolved winding number also employing the spectral localizer and demonstrate that it can topologically characterize both the $0$- and $\pi$-end states. Also, here, we find a FTAI phase with a range of disorder strengths.

As an outlook we note that using spectral localizer to extract both localizer gap and topological invariants is numerically very efficient. This makes it a very natural choice for systems in higher dimensions or for systems with real-space inhomogeneity, including disorder, where a description in reciprocal space is not available.   
One interesting system that may be explored in both two and three dimensions are higher-order topological insulators~(HOTIs)~\cite{benalcazar2017, benalcazarprb2017} and their driven counterparts Floquet HOTIs~(FHOTIs)~\cite{GhoshJPCMReview2024}. Especially for the FHOTIs, the traditional topological characterization gets rather complicated, which may open up for a spectral localizer picture~\cite{Huang2020,GhoshDynamical2022}.
Further, in this work, we only consider driven systems that belong to classes A and AIII in the AZ ten-fold class~\cite{AltlandPRB1997}. However, one should be able to generalize the spectral localizer-based invariant for all the other symmetry classes~\cite{LoringAnnPhys2015}. 

Moreover, spectral localizer-based topological invariants might be very suitable for non-Hermitian~(NH) systems~\cite{AshidaNH2020}. In fact, the spectral localizer has very recently been employed for NH systems with point gap and dislocation modes using a doubled Hamiltonian~\cite{ChadhaPRB2024}, line-gapped NH Hamiltonians~\cite{CerjanJMP2023-2}, and also in photonic systems in radiative environments in a NH description~\cite{WongnpjNanoPhoton2024}. We note particularly that in NH systems, due to the presence of skin effects, the association of boundary-protected states and topological invariants computed with periodic boundary conditions becomes more subtle~\cite{ghosh2024NHlongrange}. One way to resolve this issue is to resort to open boundary conditions (OBC)~\cite{SongNHPRL2019}. Hence, NH systems may be a perfect playground for spectral localizers, as there the topological invariants are naturally computed using OBC.

\subsection*{Acknowledgments}
We thank Lumen Eek for finding a typo in the previous version of the manuscript. We acknowledge financial support from the Knut and Alice Wallenberg Foundation through the Wallenberg Academy Fellows program KAW 2019.0309 and project grant KAW 2019.0068. Part of the computations were enabled by resources provided by the National Academic Infrastructure for Supercomputing in Sweden (NAISS), partially funded by the Swedish Research Council through grant agreement no. 2022-06725.

\appendix
\newcounter{defcounter}
\setcounter{defcounter}{0}

\begin{figure}
    \centering
    \subfigure{\includegraphics[width=0.48\textwidth]{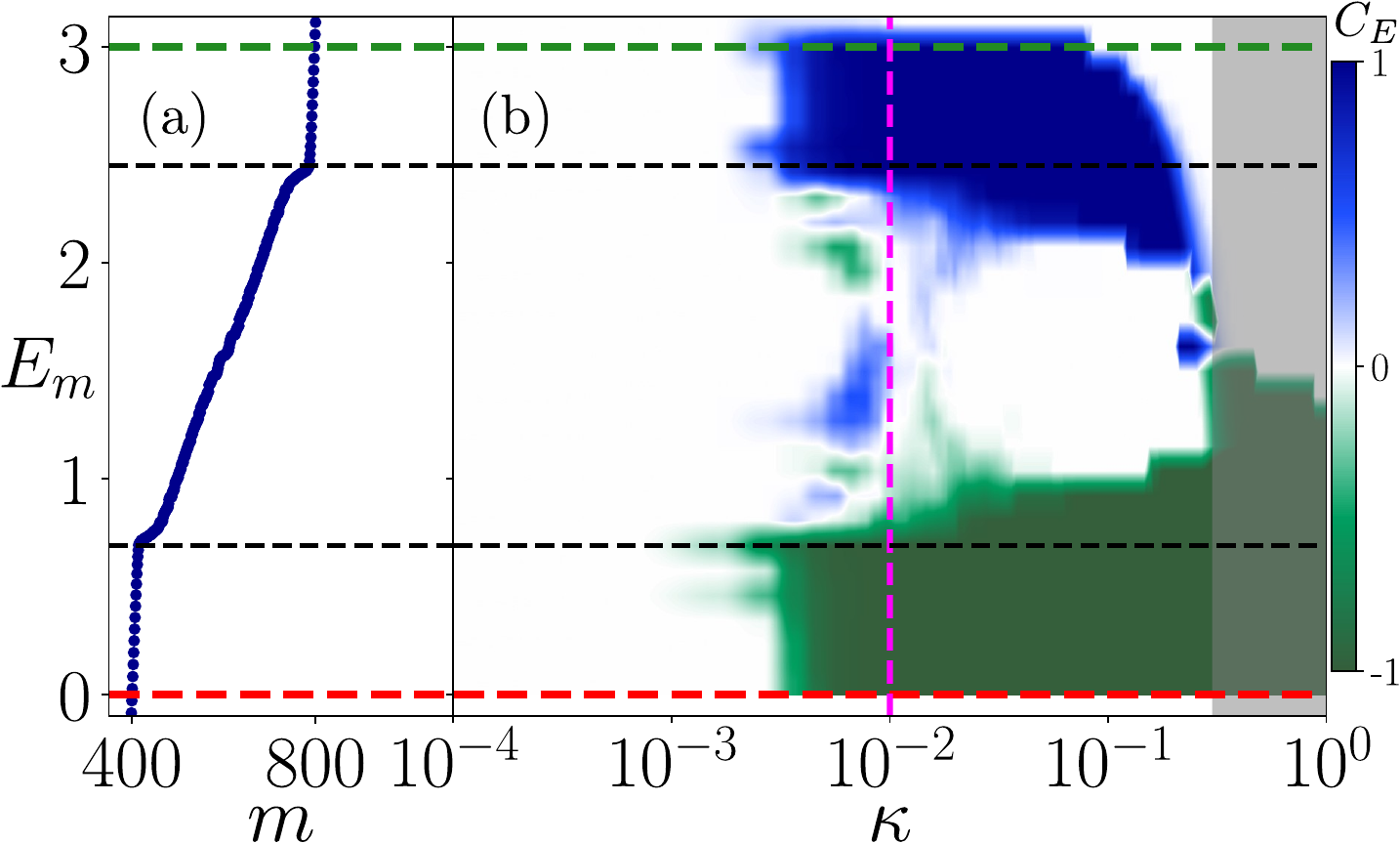}}
    \caption{(a) Positive half of the quasienergy spectrum of Fig.~\ref{Fig:2DChernAnalysis}(a). (b) Energy-resolved Chern number $C_E$ in the $\kappa \mhyphen E_m$ plane. Here, $C_E=-1$~(green) and $C_E=1$~(blue) quantifies the $0$- and $\pi$-modes, respectively. Vertical magenta dash line corresponds to the scaling constant $\kappa = 0.01$ used in the main text, while red ($E = 0$) and green ($E = \pi$) dashed lines denote the reference energy levels employed to compute $C_E$ in the main text. Region enclosed by black dashed lines approximately encloses the bulk. Gray-shaded region indicates where the computation of the invariant completely fails.
    We use the same parameters as in Fig.~\ref{Fig:2DChernAnalysis}(a).
    }
    \label{Fig:2DKappa}
\end{figure}

\section{Dependence on scaling constant $\kappa$} \label{App:A}
In the main text, we consider a fixed value of the scaling constant $\kappa = 0.01$ for all results. The scaling constant $\kappa$ maintains spectral compatibility between the position operators and the Hamiltonian of the system. For this reason, it has been argued previously that $\kappa$ can not be infinitesimally small or very high~\cite{loring2017finitevolume,loring2020spectral}. To explicitly demonstrate that our choice of $\kappa$ is appropriate, we here numerically study the behavior of the energy-resolved Chern number $C_E$ for the driven 2D Chern insulator as a function of $\kappa$. In particular, we choose parameters marked by the cyan star in Fig.~\ref{Fig:2DChern}, which are the parameters also used in Fig.~\ref{Fig:2DChernAnalysis}(a-c) and give a system harboring both $0$- and $\pi$-modes. In Fig.~\ref{Fig:2DKappa}(a), we plot the positive part of the quasienergy spectrum $E_m$ as a function of the state index $m$ and mark the quasienergies $E=0$ (red dashed line) and $E = \pi$ (green dashed line) where we extract the topological invariants.
Then, in Fig.~\ref{Fig:2DKappa}(b), we compare how $C_E$ behaves in the $\kappa \mhyphen E_m$ plane. We observe that for a very small value of $\kappa$ ($<0.001$), $C_E$ fails to capture both the $0$- and $\pi$-modes, but instead return only $C_E =0$. On the other hand, we observe that for a too high value of $\kappa$ ($\kappa>0.3$), the formalism completely fails as it returns $C_E=-2$. We do not show this value in the colorbar, but instead shade this region by gray in Fig.~\ref{Fig:2DKappa}(b).
However, for all values of $\kappa$ in between these two extremes, $C_E$ adequately captures both the $0$-modes ($C_E=-1$, green color) and $\pi$-modes ($C_E = 1$, blue color). In the main text, we choose $\kappa=0.01$, marked by the vertical magenta dashed line in Fig.~\ref{Fig:2DKappa}(b). 
Furthermore, in the main text, to compute $C_E$, we use the reference quasienergies $E=0$ and $E=\pi$ (red and green dashed lines, respectively) to be maximally within the energy gaps. Here, Fig.~\ref{Fig:2DKappa}(b) unravels another aspect of $C_E$: it is fully quantized for all in-gap energies, for all intermediate $\kappa$. Hence, it is possible to calculate the topological invariant for any in-gap energy.
We finally note that it is not very apparent from Fig.~\ref{Fig:2DKappa}(b) how $C_E$ should behave for the bulk states (within the dashed black lines), which we believe requires further analysis.

\bibliographystyle{apsrev4-2mod}
\bibliography{bibfile.bib}
\end{document}